\documentclass{article}
\usepackage[utf8]{inputenc}
\usepackage{authblk}
\usepackage{indentfirst}
\usepackage{hyperref}
\usepackage{longtable}
\usepackage{setspace}
\usepackage[margin=1in]{geometry}
\usepackage{graphicx}
\graphicspath{ {./figures/} }
\usepackage{subcaption}
\usepackage{amsmath}
\usepackage{lineno}
\usepackage{multirow}
\usepackage{color,soul}
\usepackage[version=3]{mhchem}

\usepackage[biblabel]{cite}

\title{A literature-derived dataset of migration barriers for quantifying ionic transport in battery materials}

\author[1]{Reshma Devi}
\author[1,2]{Avaneesh Balasubramanian}
\author[3,*]{Keith T. Butler}
\author[1,*]{Gopalakrishnan Sai Gautam}

\affil[1]{Department of Materials Engineering, Indian Institute of Science, Bengaluru, 560012, India}
\affil[2]{Indian Institute of Science Education and Research, Pune, 411008, India}
\affil[3]{Department of Chemistry, University College London, London WC1E 6BT, United Kingdom}
\affil[*]{Email: \href{mailto:k.t.butler@ucl.ac.uk}{k.t.butler@ucl.ac.uk}; \href{mailto:saigautamg@iisc.ac.in}{saigautamg@iisc.ac.in}}

\date{}

\begin{document}

\maketitle

\begin{abstract}
The rate performance of any electrode or solid electrolyte material used in a battery is critically dependent on the migration barrier ($E_m$) governing the motion of the intercalant ion, which is a difficult-to-estimate quantity both experimentally and computationally. The foundation for constructing and validating accurate machine learning (ML) models that are capable of predicting $E_m$, and hence accelerating the discovery of novel electrodes and solid electrolytes, lies in the availability of high-quality dataset(s) containing $E_m$. Addressing this critical requirement, we present a comprehensive dataset comprising 619 distinct literature-reported $E_m$ values calculated using density functional theory based nudged elastic band computations, across 443 compositions and 27 structural groups consisting of various compounds that have been explored as electrodes or solid electrolytes in batteries. Our dataset includes compositions that correspond to fully charged and/or discharged states of electrode materials, with intermediate compositions incorporated in select instances. Crucially, for each compound, our dataset provides structural information, including the initial and final positions of the migrating ion, along with its corresponding $E_m$ in easy-to-use .xlsx and JSON formats. We envision our dataset to be a highly useful resource for the scientific community, facilitating the development of advanced ML models that can predict $E_m$ precisely and accelerate materials discovery.
\end{abstract}


\section{Background and Summary}
Ionic conductivity ($\sigma$) is one of the most important properties that is used to characterize materials used for electrochemical applications, such as a battery electrode or an electrolyte.\cite{rong2015materials,euchner2020stability,park2010review, bachman2016inorganic} Typically, ionic conduction is a thermally activated process defined by the Nernst-Einstein equation as, 
\begin{equation}
\sigma = \frac{q^2 x D(x)}{k_B T}    
\end{equation}
Where $q$ and $x$ are the charge and concentration of the intercalant (or the electroactive ion), respectively. $D(x)$ is the diffusion coefficient of the intercalant that varies with $x$, $T$ is the temperature and $k_B$ is the Boltzmann constant. $D(x)$ relates the diffusive flux ($J$) and the concentration gradient ($\nabla x$ of the intercalating species via the Fick’s first law ($J=-D(x)\nabla x$).\cite{fick1855v} Further, $D(x)$ can be written as, 
\begin{equation}\label{eq:dx}
D(x) = D_J(x) \theta(x) 
\end{equation}
$D_J$ is the jump diffusion coefficient, which captures all the cross correlations among the individual atomic migrations and $\theta$ is the thermodynamic factor that captures the non-ideality of the solid solution (i.e., the interactions between the migrating ions and the host framework). $\theta$ is defined as $\theta = \frac{\partial (\mu / k_B T)}{\partial \ln x}$,  where $\mu$ is the chemical potential of the migrating ion. In solid electrodes and electrolytes, $x$ is typically the site fraction of the migrating ion. For an ideal solid solution where each ionic hop has an identical hop frequency that is independent of the local concentration/configuration, $D(x)$ becomes, 
\begin{equation} \label{eq:diffusion_coefficient}
D = fga^2 \nu \exp\left(-\frac{E_m}{k_B T}\right)
\end{equation}
$g$ is the geometric factor that determines how the diffusion channels are connected, $f$ is the correlation factor, $a$ and $\nu$, are the hop distance and vibrational prefactor, respectively, and $E_m$ is the activation energy of migration. Ion transport within a crystalline lattice occurs through ionic migration events, where an ion moves from its original or interstitial site in a lattice to a neighboring vacant site, via a transition state. The migration process is influenced by the energy landscape encountered by the ion during its movement, with the $E_m$ playing a crucial role in determining the ease of ionic mobility and, consequently, the material's $\sigma$. 

Extensive research has focused on enhancing ionic conductivity by minimizing $E_m$, as this directly improves the rate capabilities of battery systems. Previous studies have shown the underlying host structure to play a vital role in influencing $D$, such as the presence of interconnected prismatic sites leading to improved Na$^+$ mobility in P2-type layered structures.\cite{clement2017direct} In compositions like LiNiO$_2$, Li off-stoichiometry leading to Ni$^{2+}$ ions in the Li layers obstructing diffusion pathway can effect the Li-ion conductivity significantly.\cite{bianchini2019there} Indeed, dopants that stabilize the layered structure, such as Ti$^{4+}$ have been used to improve Na$^+$ mobility.\cite{vassilaras2017communication} In the case of phosphates, nuclear magnetic resonance (NMR)\cite{grey2004nmr,verhoeven2001lithium} studies reveal that intercalant diffusivity is not governed by a single, uniform barrier but by a distribution of local energy barriers that are dictated by the arrangement of neighboring transition metal cations.\cite{strobridge2014characterising} Additionally, subtle electrostatic distortions that screen electrostatic interactions between the intercalant and the anion framework have been shown to improve intercalant mobility in polyanionic structures.\cite{bianchini2014na3v2} 

Galvanostatic intermittent titration technique (GITT)\cite{kang2021galvanostatic,tang2011kinetic}  measurements in Mn and Fe rich disordered rocksalt structures have revealed the importance of Li-exccess compositions, particle size, and the underlying redox process as some of the important factors that affect the intercalant diffusivity.\cite{lee2015new,lee2021determining,fong2024redox} Bonnick et al. illustrated the influence of poor electronic conductivity resulting in strong electrostatic interactions within thiospinel lattices (e.g., MgZr$_2$S$_4$) resulting in a  reduction of ionic diffusivity.\cite{bonnick2018insights} In summary, various structural and chemical modifications have been explored across different types of intercalation compounds, including layered, spinel, olivine, polyanionic, and other frameworks, to enhance ionic conductivity,\cite{rong2015materials,lu2021searching,wang2015design,canepa2017high} with some approaches using targeted machine learning (ML) techniques as well.\cite{jalem2014efficient,jalem2018bayesian,sendek2017holistic} However, developing universal optimization strategies across a wide range of intercalation systems remains challenging due to the interplay between structure, composition, and other factors besides the lack of a robust $E_m$ dataset that spans a diverse range of materials.

In general, estimating diffusivities or $E_m$ using experimental techniques like variable temperature NMR, GITT, and electrochemical impedance spectroscopy (EIS),\cite{barsukov2012electrochemical, itagaki2005licoo2} are either experimentally challenging or resource intensive. This is due to the extremely short time scales (10$^{-12}$ s) or small length scales ($\sim$few \AA{}) of the elementary process of ionic hopping, influence of surface and structural chemistry of electrodes/electrolytes on the measurement, variations in sample preparation and measurement conditions resulting in differences in interfacial formation, bulk stoichiometry and defects, and specific equipment requirements (e.g., the need for inert ion-blocking electrodes in EIS). Thus, experimental NMR/GITT/EIS data documenting $E_m$ is unavailable for a wide range of materials.

Computational methodologies to estimate $E_m$ have gained prominence, since calculated $E_m$ can be used as a screening metric within high-throughput workflows before experimental validation. Computational techniques include empirical approaches such as bond valence sum (BVS)\cite{adams2006bond,brown2009recent} analysis and first principles simulations based on density functional theory (DFT)\cite{hohenberg1964inhomogeneous,kohn1965self} or molecular dynamics (MD).\cite{frenkel2002understanding,mo2012first,car1985unified} BVS analysis, though computationally swift, has accuracy limitations as it relies on an ionic bond model, making it more suitable close-packed lattices with highly electronegative anions.\cite{meutzner2019computational,nestler2019towards} Nudged elastic band (NEB\cite{jonsson1998nudged,henkelman2000climbing}) calculations based on DFT can provide accurate $E_m$ by modeling the ionic migration path using intermediate images that are connected by fictitious spring forces and subsequently relaxing the images to identify the saddle point that corresponds to the transition state. However, the DFT-NEB approach is computationally intensive for large systems ($>$100 atoms), and its accuracy/convergence can depend on the chosen exchange-correlation (XC) functional within DFT.\cite{devi2022effect} Classical MD and ab-initio MD techniques can directly estimate $D(x)$ at multiple $T$, thus yielding $E_m$ from Equation~{\ref{eq:diffusion_coefficient}}, but are computationally demanding due to the time and length scales that need to be captured.\cite{he2018statistical} 

Some strategies have been explored to reduce the computational costs and constraints, while retaining the accuracy, of both the DFT-NEB and ab-initio MD approaches. For example, the `pathfinder' approach in conjunction with the `ApproxNEB' scheme\cite{rong2016efficient} aims to reduce the computational constraints of DFT-NEB by mitigating convergence issues via selection of a `better' initial migration path for calculation. However, the scheme still requires performing a full DFT-NEB calculation and is prone to the underlying constraints of the DFT-NEB approach. Another pathway is integrating ab-initio MD simulations with machine learned interatomic potentials (MLIPs), where the MLIPs can theoretically provide higher computational speeds with the accuracy comparable to DFT.\cite{deringer2019machine} While several MLIP frameworks that are accurate remain chemistry-specific (i.e., there are high computational costs associated with training the MLIPs),\cite{choyal2024constructing, gartner2020signatures,achar2021efficiently} the foundational or universal MLIPs\cite{batatia2022design, batatia2023foundation,rhodes2025orb,deng2023chgnet,chen2022universal,kim_sevennet_mf_2024} have not been tested rigorously on $D(x)$ or $E_m$ predictions, yet. More importantly, we need datasets of $E_m$ that are available over a wide-range of chemistries and structures to be able to test universal MLIPs in their utility in predicting $E_m$ and/or build ML models that are tailored to accurately predict $E_m$ that can be used for screening through materials.

In this work, we present a literature-based curated dataset of 619 DFT-NEB-derived $E_m$ values across various compounds that have been studied as electrodes or solid electrolytes in lithium, sodium, potassium, and multivalent ion based battery systems. Our dataset includes fully charged and discharged states of electrode materials, with intermediate (non-stoichiometric) compositions considered in 30 cases. Additionally, we provide structural information for each compound, including the initial and final positions of the migrating ion, along with its corresponding energy barrier, which can be used in the construction of ML models that require structural inputs, such as graph-based models leveraging transfer learning.{\cite{devi2024optimal}} Our dataset includes a total of 275 distinct entries contributed by 99 systems exhibiting multiple migration pathways. We envision our dataset to be a powerful resource for the scientific research and industrial communities, enabling the development of robust ML models and MLIPs that can eventually accelerate materials discovery for batteries and other applications.

\section{Methods}
We conducted a thorough manual review of battery research articles published over the past two decades to compile computationally estimated $E_m$ values. To ensure consistency and reliability, we focused exclusively on DFT-NEB calculated $E_m$ values, as this method strikes a balance between accuracy and availability in the literature compared to other computational approaches. We note that the generalized gradient approximation (GGA\cite{perdew1996generalized}) is the most widely used XC functional among DFT-NEB calculations. Other functionals that are commonly used include Hubbard \textit{U}\cite{anisimov1991band} corrected GGA (i.e., GGA+\textit{U}\cite{wangoxidation2006}), local density approximation (LDA\cite{jones1989density}), strongly constrained and appropriately normed (SCAN\cite{sun2015strongly}) and its Hubbard \textit{U}-corrected variant (SCAN+\textit{U}\cite{sai2018evaluating}). Since GGA (and GGA+\textit{U}) is computationally less intensive than SCAN/SCAN+\textit{U} and gives reasonably accurate $E_m$ estimations,\cite{devi2022effect}  GGA (or GGA+\textit{U}) is the preferred choice for DFT-NEB calculations. Our dataset reflects this preference, with 88.05\% of the collected $E_m$ values calculated using GGA, followed by GGA+\textit{U} (7.27\%), SCAN (3.07\%) and LDA (1.61\%). In cases where multiple XC functionals were used to calculate $E_m$ within the same or different research articles, we prioritised GGA-calculated values to maintain consistency with the majority of the dataset. Note that for structures where the GGA/GGA+\textit{U}-calculated $E_m$ was not available, we included the $E_m$ value as calculated with the different XC functional. Thus, the $E_m$ barriers of non-GGA functionals have been used as is, and we have not done any re-calculation of such systems to ensure that all $E_m$ have been calculated at the same level of theory.

We considered a research article for inclusion in our dataset only if it satisfied specific criteria ensuring the reliability and completeness of the reported DFT-NEB $E_m$ values, i.e., we included studies that provided a comprehensive methodology for $E_m$ calculations. Details on the methods we looked for in articles included the XC functional used, the number of intermediate images used to describe the migration pathway, and the supercell size employed in the NEB calculations. Articles that lacked the aforementioned details were excluded to maintain data consistency and accuracy. Structural information was another key criterion for selecting article, where we only considered studies that provided explicit description of the parent structure(s), including the space group(s) and lattice parameters. In case of multiple articles reporting different $E_m$ values for a given structure, we included the paper (and the corresponding $E_m$) with the most details on the structure and the methods. If the structural parameters were missing and could not be retrieved from related works or structural repositories, we excluded the corresponding study from the dataset. For materials exhibiting multiple migration pathways, we ensured that the reported $E_m$ were appropriately distinguished for each path via clear descriptions of the corresponding pathways. In cases where $E_m$ values were not explicitly stated in text but were presented through minimum energy pathway (MEP) plots or other visualizations, only articles with clear, well-labeled plots featuring correct units and axis scales were selected to ensure accurate digital data extraction. 

\subsection*{Workflow}
\begin{figure}[h!]
\centering
\includegraphics[width=\linewidth]{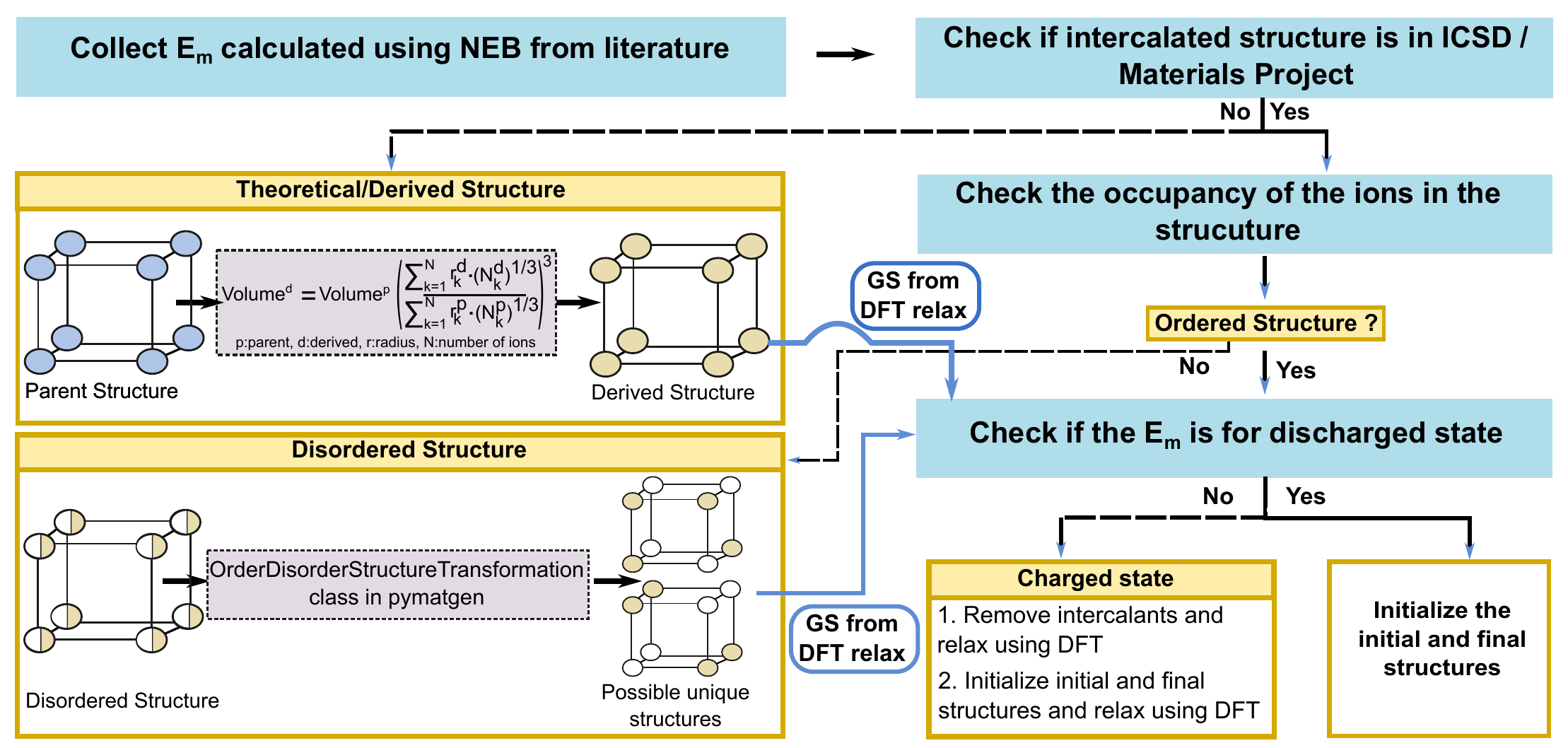}
\caption{Flowchart illustrating the structural data generation process for each data point in the database. GS refers to ground state. Relax refers to the structural relaxation calculation done with DFT.}
\label{fig:flowchart}
\end{figure}

Figure~{\ref{fig:flowchart}} presents an overview of the workflow for generating the structural information for our dataset. We analyzed the collected structural information for each datapoint, and ensured that the target structures were download from either the inorganic crystal structure database (ICSD\cite{hellenbrandt2004inorganic}) or the materials project (MP\cite{jain2013commentary}). If a target structure was available in both ICSD and MP, we downloaded the computationally relaxed structure from the MP. For all electrode materials, where possible, we downloaded the structure of the discharged composition (i.e., structures with high concentrations of intercalant ions, relevant for electrode materials) preferably over the charged composition.

If the target structure was not available in either ICSD or MP, we searched for a parent structure that shared the same space group, migrating ion concentration, and site occupancies, but containing ions that are different from the target structure in both ICSD and MP. The parent structure was then used as a template, where we substituted the occupant ions with the target ions and used the reference lattice scaling (RLS\cite{chu2018predicting}) scheme, as implemented by the RLSVolumePredictor class in the pymatgen package,\cite{ong2013python} to obtain a target structure with scaled lattice parameters and the right composition. If target structures were available but contained sites with partial occupancies, we enumerated all possible symmetrically unique configurations that satisfied the target stoichiometry of the reported structure by using the OrderDisorderStructureTransformation class in pymatgen. For all the RLS-generated and/or enumerated structures that we obtained, we performed structure relaxations with DFT to obtain the ground state (i.e., lowest energy) configuration.

For all structure relaxations with DFT, we used the GGA XC functional as available in the Vienna ab initio simulation package (VASP, version 6.1.2).\cite{kresse1993ab,kresse1996efficient} The effects of core electrons were described via projector augmented wave potentials.\cite{kresse1999ultrasoft} We relaxed the lattice vectors, cell volume, and ionic positions of all input structures, without preserving any symmetry, until the total energy and atomic forces were below 0.01 meV and $|0.03|$ eV/{\AA}, respectively. We used a $\Gamma$-centered $k$-point mesh with a density of at least 48 $k$-points per \AA{}$^{-1}$ for sampling the irreducible Brillouin zone, a kinetic energy cutoff of 520 eV for describing the plane-wave basis set, and a Gaussian smearing of width 0.05 eV to intergrate the Fermi surface. For select structures that exhibited convergence difficulties with GGA, we performed the structural relaxations with GGA+\textit{U}, utilizing the optimized values reported in Ref.~\citenum{wangoxidation2006} All our structures are compatible with the structure relaxation calculation settings of MP (version 2020).

Upon obtaining the ground states for all systems in our dataset, we reviewed the migration path information reported in the corresponding research articles and initialized the initial and final configurations of the migration path within each structure. Note that the initial configuration represents the migrating ion occupying the starting site while leaving the destination site vacant, while the final configuration depicts the inverse arrangement. Thus, we have assumed that all migration events for all structures considered in our dataset occur via a vacancy-based mechanism and not an interstitial-based mechanism. All the solid electrolyte materials in our dataset are stoichiometric and ordered compositions, hence we treated them as equivalent to discharged compositions of electrode materials in our workflow. 

For electrode structures where the $E_m$ was reported for a charged composition (i.e., the dilute ion limit), we removed the intercalant ions from the ground state discharged structure and subsequently relaxed the structure using DFT. The initial and final configurations were then defined in this relaxed charged structure and were DFT-relaxed again to obtain their true ground state descriptions. In the case of intermediate intercalant compositions, all symmetrically distinct positional configurations corresponding to the composition were enumerated, followed by DFT relaxation to identify the ground state, and the corresponding pathway initializations were carried out in the ground state. For generating all initial and final configurations, we selected appropriate supercell sizes to ensure that the migrating ion does not experience spurious interactions by maintaining a minimum distance of at least 8 \AA{} with its periodic images. In structures with large unit cells, such as NaSICONs (sodium superionic conductors), weberites, and oxyfluorides, we did not generate supercells to reduce computational costs.

\section{Data records}

We report computationally calculated $E_m$ of 619 systems that have been explored as battery materials along with their structural information for each possible ionic migration event. The data can be easily downloaded in the form of a JSON file from our \href{https://github.com/sai-mat-group/migration-barrier-dataset}{GitHub} repository.

\subsection*{File format}

Each datapoint in the dataset is associated with specific tags that provide essential information, as summarized in Table~{\ref{tab:params}}. For each datapoint, we include its composition, crystal class, space group, unique system identifier, and a JSON ID that differentiates each datapoint within the database and allows easier access to different migration paths within a given structure. We assign the system name for each datapoint using a standardized format: \textit{reduced\_chemical\_formula} + ‘\_’ + \textit{path\_number}. For instance, MgCoSiO{\textsubscript{4}} has two possible migration paths for Mg$^{2+}$ diffusion, so we represent each path as \textit{MgCoSiO4\_1} and \textit{MgCoSiO4\_2}. However, if a composition has only one active migration path, it is identified using only the reduced chemical formula. Parentheses and subscripts are omitted in the system name generation. 

When multiple polymorphs of the same composition exist, the system name is modified to include the space group: \textit{reduced\_chemical\_formula} + ‘\_’ +\textit{ space\_group} + ‘\_’ + \textit{path\_number}.  Additionally, for layered structures where both monovalent and divalent hops are considered,\cite{van2001first} we treated both hops as distinct migration paths. Charged state structures are labeled using the format: \textit{charged\_state\_reduced\_chemical\_formula} + \textit{\_} + \textit{intercalant}, allowing clear differentiation from the discharged state configurations. In order to be compatible with the notations used in the original papers, certain prefixes, such as O3:O3-layered structure, O:olivines, M:maricite, d:$\delta$, e:$\epsilon$, g:$\gamma$, b:$\beta$, a:$\alpha$, and a1:$\alpha$1, were added to the corresponding system names to ease the identification. We did not modify our notation for intermediate compositions as compared to the discharged compositions.

The dataset provides structural information for both the initial and final configurations of each migration path in the ‘POSCAR’ format, which is compatible with pymatgen and the atomic simulation environment (ASE\cite{larsen2017atomic}) packages, enabling easy conversion into other structural representations. Each datapoint also includes a "bibtex" tag, which contains the citation details of the article from which the $E_m$ values and the migration path information were sourced. Additionally, the XC functional used to calculate the $E_m$ in the respective study is provided under the "XC" tag for reference. 

\begingroup
    \begin{table}[ht!]
    \centering
        \begin{tabular}{|c|c|c|}
        \hline
        \textbf{S.No} &\textbf{Tags} &\textbf{Description} \\
        \hline
        \centering 1 & jid  &  Unique JSON ID  \\
        \hline
        \centering 2 &  structure\_ini   &  Initial configuration of the migration path  \\
        \hline
        \centering 3 & structure\_fin   &  Final configuration of the migration path  \\
        \hline
        \centering 4 & target    & Reported $E_m$   \\
        \hline
        \centering 5 & formula  &  Chemical composition \\
        \hline
        \centering 6 & crystal\_class   &  The crystal class  \\
        \hline
        \centering 7 &  sys\_name    &  A unique system name   \\
        \hline
        \centering 8 &  space\_group     &  Space group   \\
        \hline
        \centering 9 &  XC    &  XC functional used for the NEB calculation in literature  \\
        \hline
        \centering 10 &  bibtex  & Citation of the research article in 'bibtex' format  \\
        \hline
        \end{tabular}
        \caption{Description of each tag associated with the datapoints in the $E_m$ dataset. }
        \label{tab:params}
    \end{table} 
\endgroup

\subsection*{Data distribution}
\begin{figure}[h!]
\centering
\includegraphics[width=\linewidth]{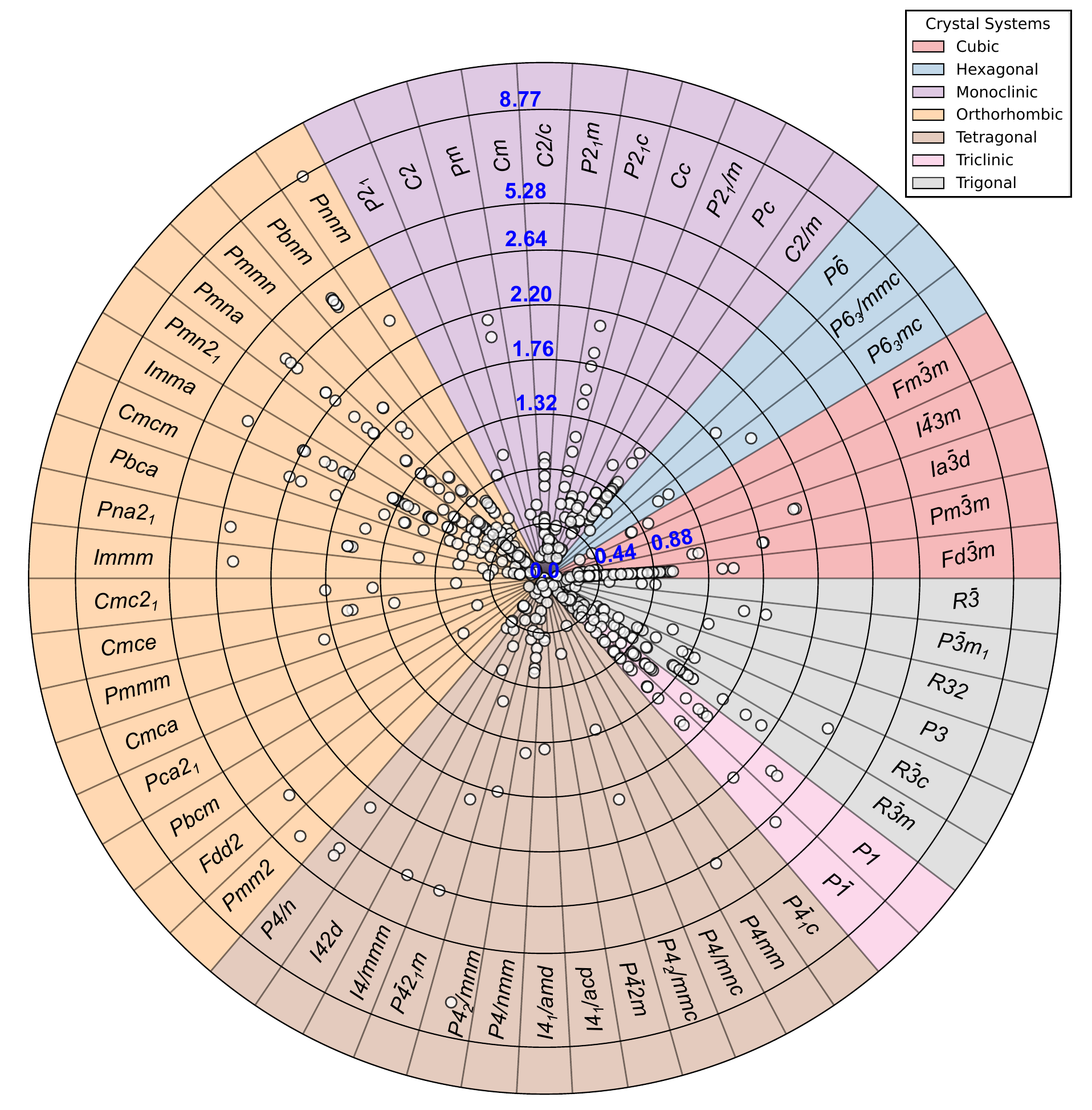}
\caption{Contour plot illustrating the distribution of the $E_m$ dataset over different space groups from each of the seven crystal systems. Individual colored sectors represent individual crystal systems, with space groups indicated by text notations. White circles indicate invidual data points. Concentric circles correspond to different E$_m$ values (in eV), as highlighted by the blue text notations.}
\label{fig:SG}
\end{figure}

The distribution of $E_m$ across the seven crystal systems, and the corresponding space groups within each crystal system, in our dataset is illustrated as a contour plot in Figure~{\ref{fig:SG}}. Each crystal system is visually distinguished using color-coded sectors in Figure~{\ref{fig:SG}}, with the solid concentric rings representing different $E_m$ values (in eV). Overall, our dataset includes $E_m$ values spanning 58 space groups and ranging from 0.03 eV to 8.77 eV. Out of the 619 datapoints, 528 are electrodes and 91 are electrolytes. The lowest $E_m$ (0.03 eV) corresponds to charged-LiTiO$_2$ (\textit{P42/mnm}), while the highest (8.77 eV) is observed in discharged-LiRuO$_2$ (\textit{Pnnm)}, respectively. Notably, both LiTiO$_2$ and LiRuO$_2$ adopt the rutile-type structure and consequently exhibit the widest range of $E_m$ values in the dataset. In contrast, space groups $Ia\overline{3}d$ and \textit{I41/acd}, which contain two and three datapoints, respectively, show the narrowest range of $E_m$ values. 

The space group $Fd\overline{3}m$, corresponding to cubic spinels, has the highest number of datapoints, contributing 94 entries to the dataset. It is followed by the \textit{Pmna} space group (72 datapoints), belonging to the orthorhombic system, and the \textit{P1} space group (39 datapoints) from the triclinic system. Additionally, 15 space groups contribute only a single datapoint each. Among the different crystal systems, orthorhombic accounts for the highest (205) number of datapoints, whereas hexagonal (6) contributes the least. The dataset exhibits a mean $E_m$ value of 0.848 eV with a standard deviation of 0.824 eV, indicating a highly skewed distribution. A majority (73.4\%) of the $E_m$ values are below 1 eV, while 19.4\% fall within the 1–2 eV range, and 7.2\% exceed 2 eV. 71.4\% and 23.6\% of the entries are contributed by discharged (high intercalant content) and charged (low intercalant content) state structures respectively, corresponding to 106 distinct charged and discharged pairs. Intermediate intercalant compositions correspond to 5\% of the dateset. 

\begin{figure}[h!]
\centering
\includegraphics[width=\linewidth]{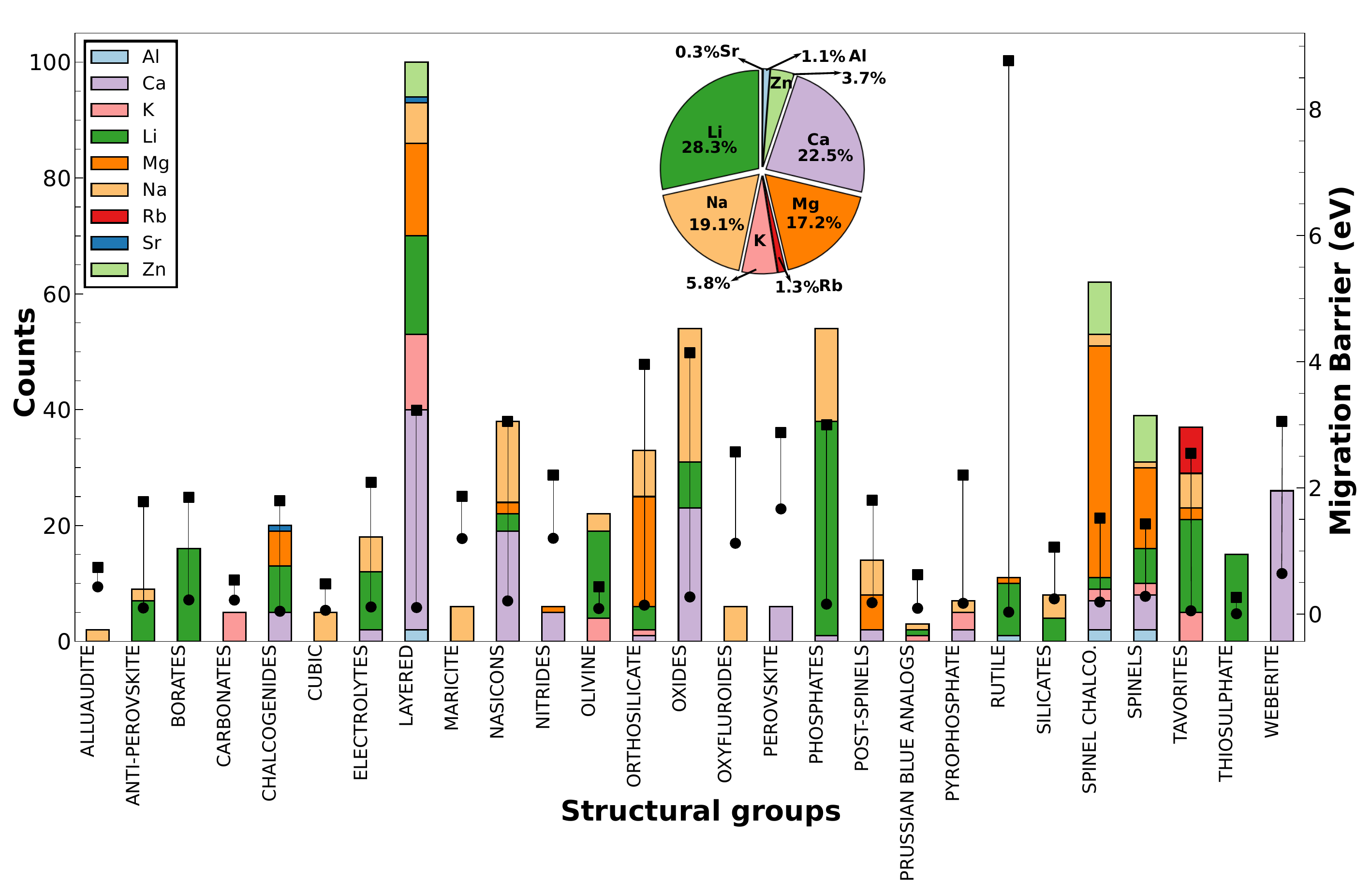}
\caption{Illustration of the $E_m$ distribution within each of the different structural groups. The \emph{y} axis on the left and right represent the number of datapoints and $E_m$ values (in eV), respectively. Stacked bar charts correspond to the counts within each structural group, with the colors indicating the split across various intercalants. The black vertical lines represent the range of $E_m$ values for a given structural group with the squares and circles representing the maxima and minima, respectively. The inset shows a pie-chart with the contributions from each intercalant (i.e., Al, Ca, K, Li, Mg, Na, Rb, Sr, and Zn, as represented by the different colors) to the total dataset.}
\label{fig:bar}
\end{figure}

Figure~{\ref{fig:bar}} presents a bar chart that illustrates the number of datapoints and the range of $E_m$ values across 27 different structural groups (e.g., spinels, olivines, NaSICONs, etc.). Each bar is stacked to represent contributions from nine different intercalating ions, with the stack length indicating the number of datapoints associated with each ion. The inset pie chart provides a breakdown of the percentage contribution of each intercalating ion to the overall dataset. Additionally, the solid square and circle markers, connected by a vertical line, denote the maximum and minimum $E_m$ values observed within each structural group, thus representing the range of $E_m$. 

Lithium-based intercalant compounds constitute 28.27\% of the dataset, making Li the most prevalent intercalating ion, which is expected given the extensive research done on Li-based electrodes and solid electrolytes. Li is followed by calcium (Ca), sodium (Na), and magnesium (Mg) in terms of contribution. The least represented intercalants are strontium (Sr), aluminum (Al), and rubidium (Rb), with only 2, 7, and 8 datapoints, respectively. 17 out of the 27 structural groups include compounds intercalating Li, whereas Ca-based compounds are primarily found in layered structures, NaSICONs, oxides, and weberites. Na-based compounds are more widely distributed than Li, appearing in 19 of the 27 structural groups, while Mg-based compounds are predominantly found in spinel chalcogenides, layered structures, and orthosilicates. Layered structures contribute the highest number of datapoints, with 98 entries, followed by spinel chalcogenides, phosphates, and oxides. Other structural groups, such as alluaudites, Prussian blue analogs, and carbonates, are also represented but in significantly smaller numbers. The highest and lowest $E_m$ ranges among the structural groups are observed in rutiles and thiosulphates, re

\section{Technical validation}
A benchmarking of DFT-NEB E$_m$ (using different XC functionals) against experimentally reported values has been performed in our previous work.\cite{devi2022effect} To estimate $E_m$, we fully relaxed the endpoint geometries representing the initial and final states of migration using DFT. Subsequently, the MEP was initialized by linearly interpolating both atomic positions and lattice vectors to create seven intermediate images between the endpoints, with a spring force constant of 5 eV/Å$^2$ maintained between adjacent images. The images constituting the NEB were optimized along the reaction coordinate using the limited-memory Broyden-Fletcher-Goldfarb-Shanno (L-BFGS)\cite{nocedal1980updating} method until the force component perpendicular to the elastic band fell below |0.05| eV/Å. The total energy of each image was also converged to within 0.01 meV. All $E_m$ values were determined assuming a vacancy-mediated mechanism in the dilute-vacancy limit. 

On comparing the calculated $E_m$ against experimental values, we observed the SCAN functional to exhibit a higher accuracy on an average relative to other XC functionals. Albeit the higher accuracy of SCAN is counterbalanced by increased computational expense and potential convergence problems. Also, we found GGA to be a suitable alternative for quick and qualitative $E_m$ predictions. 

Approximately 15.8\% of the dataset has been calculated from scratch using the GGA or SCAN XC functional (depending on the computational feasibility for the respective system), using the above described methodology and have also been reported in other works. \cite{tekliye2022exploration,tekliye2024fluoride,deb2024exploration,gautam2015first,gautam2016impact,devi2022effect}. The remaining 84.2\% of the dataset has been collected from various literature sources that report DFT-NEB methodologies similar to the above description. In certain cases, articles that used fewer number of images (3 or 5) were also considered if a fully convereged MEP was reported.

\section{Usage notes}
We present a literature-curated DFT-NEB-calculated dataset comprising 619 distinct $E_m$ values over 443 chemistries and 27 distinct structural groups, spanning a diverse set of electrode and solid electrolyte materials studied for battery applications. This dataset, which includes structural information and calculated $E_m$ values for each system, is provided in both .xlsx and JSON formats for easier and direct data extraction within our \href{https://github.com/sai-mat-group/migration-barrier-dataset}{GitHub} repository.

Our dataset can be effectively utilized to construct ML models for $E_m$ estimation, using either structural, compositional, or combined inputs. Additionally, it is suitable for fine-tuning pre-trained foundational models and for benchmarking the performance of various universal MLIPs on $E_m$ prediction tasks.

\section{Data availability}
The dataset developed as part of this work is available freely online at our \href{https://github.com/sai-mat-group/migration-barrier-dataset}{GitHub} repository.

\section{Code availability}
There are no specific codes or scripts developed in this work. Licenses for using the VASP code, which was employed for the DFT calculations done in this work, are available at \href{https://vasp.at/}{https://vasp.at/}. The pymatgen package can be downloaded freely from \href{https://pymatgen.org/}{https://pymatgen.org/}.

\section{Acknowledgements}
G.S.G. and K.T.B. would like to acknowledge financial support from the Royal Society under grant number IES$\backslash$R3$\backslash$223036, and the United Kingdom Research and Innovation (UKRI) Engineering and Physical Sciences Research Council (EPSRC), under projects EP/Y000552/1 and EP/Y014405/1. G.S.G. acknowledges financial support from the Science and Engineering Research Board (SERB) of the Department of Science and Technology, Government of India, under sanction number IPA/2021/000007. R.D. thanks the Ministry of Human Resource Development, Government of India, for financial assistance. R.D. and G.S.G. acknowledge the computational resources provided by the Supercomputer Education and Research Centre, IISc, for enabling some of the calculations showcased in this work. We acknowledge National Supercomputing Mission (NSM) for providing computing resources of ‘Param Utkarsh’ at CDAC Knowledge Park, Bengaluru. PARAM Utkarsh is implemented by CDAC and supported by the Ministry of Electronics and Information Technology (MeitY) and Department of Science and Technology (DST), Government of India. Via our membership of the UK's HEC Materials Chemistry Consortium, which is funded by EPSRC (EP/X035859/1), this work used the ARCHER2 UK National Supercomputing Service (\href{http://www.archer2.ac.uk}{http://www.archer2.ac.uk}).

\bibliographystyle{unsrt}
\bibliography{sample}

\begin{thebibliography}{10}

\bibitem{rong2015materials}
Ziqin Rong, Rahul Malik, Pieremanuele Canepa, Gopalakrishnan Sai~Gautam, Miao
  Liu, Anubhav Jain, Kristin Persson, and Gerbrand Ceder.
\newblock Materials design rules for multivalent ion mobility in intercalation
  structures.
\newblock {\em Chemistry of Materials}, 27(17):6016--6021, 2015.

\bibitem{euchner2020stability}
Holger Euchner, Jin~Hyun Chang, and Axel Gross.
\newblock On stability and kinetics of li-rich transition metal oxides and
  oxyfluorides.
\newblock {\em Journal of Materials Chemistry A}, 8(16):7956--7967, 2020.

\bibitem{park2010review}
Myounggu Park, Xiangchun Zhang, Myoungdo Chung, Gregory~B Less, and Ann~Marie
  Sastry.
\newblock A review of conduction phenomena in li-ion batteries.
\newblock {\em Journal of power sources}, 195(24):7904--7929, 2010.

\bibitem{bachman2016inorganic}
John~Christopher Bachman, Sokseiha Muy, Alexis Grimaud, Hao-Hsun Chang, Nir
  Pour, Simon~F Lux, Odysseas Paschos, Filippo Maglia, Saskia Lupart, Peter
  Lamp, et~al.
\newblock Inorganic solid-state electrolytes for lithium batteries: mechanisms
  and properties governing ion conduction.
\newblock {\em Chemical reviews}, 116(1):140--162, 2016.

\bibitem{fick1855v}
Adolph Fick.
\newblock V. on liquid diffusion.
\newblock {\em The London, Edinburgh, and Dublin Philosophical Magazine and
  Journal of Science}, 10(63):30--39, 1855.

\bibitem{clement2017direct}
RJ~Cl{\'e}ment, J~Xu, DS~Middlemiss, J~Alvarado, C~Ma, YS~Meng, and CP~Grey.
\newblock Direct evidence for high na+ mobility and high voltage structural
  processes in p2-nax[liynizmn1-y-z]o2 (x{,} y{,} z ? 1) cathodes from
  solid-state nmr and dft calculations.
\newblock {\em Journal of Materials Chemistry A}, 5(8):4129--4143, 2017.

\bibitem{bianchini2019there}
Matteo Bianchini, Maria Roca-Ayats, Pascal Hartmann, Torsten Brezesinski, and
  J{\"u}rgen Janek.
\newblock There and back again—the journey of linio2 as a cathode active
  material.
\newblock {\em Angewandte Chemie International Edition}, 58(31):10434--10458,
  2019.

\bibitem{vassilaras2017communication}
Plousia Vassilaras, Stephen~T Dacek, Haegyeom Kim, Timothy~T Fister, Soojeong
  Kim, Gerbrand Ceder, and Jae~Chul Kim.
\newblock Communication—o3-type layered oxide with a quaternary transition
  metal composition for na-ion battery cathodes: Nati0. 25fe0. 25co0. 25ni0.
  25o2.
\newblock {\em Journal of The Electrochemical Society}, 164(14):A3484, 2017.

\bibitem{grey2004nmr}
Clare~P Grey and Nicolas Dupr{\'e}.
\newblock Nmr studies of cathode materials for lithium-ion rechargeable
  batteries.
\newblock {\em Chemical reviews}, 104(10):4493--4512, 2004.

\bibitem{verhoeven2001lithium}
VWJ Verhoeven, IM~De~Schepper, G~Nachtegaal, APM Kentgens, EM~Kelder,
  J~Schoonman, and FM~Mulder.
\newblock Lithium dynamics in limn 2 o 4 probed directly by two-dimensional 7
  li nmr.
\newblock {\em Physical review letters}, 86(19):4314, 2001.

\bibitem{strobridge2014characterising}
Fiona~C Strobridge, Derek~S Middlemiss, Andrew~J Pell, Michal Leskes,
  Rapha{\"e}le~J Cl{\'e}ment, Fr{\'e}d{\'e}rique Pourpoint, Zhouguang Lu,
  John~V Hanna, Guido Pintacuda, Lyndon Emsley, et~al.
\newblock Characterising local environments in high energy density li-ion
  battery cathodes: a combined nmr and first principles study of life x co 1- x
  po 4.
\newblock {\em Journal of Materials Chemistry A}, 2(30):11948--11957, 2014.

\bibitem{bianchini2014na3v2}
Mateos Bianchini, Nicolas Brisset, Fran{\c{c}}ois Fauth, Fran{\c{c}}ois Weill,
  Erik Elkaim, Emmanuelle Suard, Christian Masquelier, and Laurence Croguennec.
\newblock Na3v2 (po4) 2f3 revisited: a high-resolution diffraction study.
\newblock {\em Chemistry of Materials}, 26(14):4238--4247, 2014.

\bibitem{kang2021galvanostatic}
Stephen~Dongmin Kang and William~C Chueh.
\newblock Galvanostatic intermittent titration technique reinvented: Part i. a
  critical review.
\newblock {\em Journal of The Electrochemical Society}, 168(12):120504, 2021.

\bibitem{tang2011kinetic}
Kun Tang, Xiqian Yu, Jinpeng Sun, Hong Li, and Xuejie Huang.
\newblock Kinetic analysis on lifepo4 thin films by cv, gitt, and eis.
\newblock {\em Electrochimica Acta}, 56(13):4869--4875, 2011.

\bibitem{lee2015new}
Jinhyuk Lee, Dong-Hwa Seo, Mahalingam Balasubramanian, Nancy Twu, Xin Li, and
  Gerbrand Ceder.
\newblock A new class of high capacity cation-disordered oxides for
  rechargeable lithium batteries: Li--ni--ti--mo oxides.
\newblock {\em Energy \& Environmental Science}, 8(11):3255--3265, 2015.

\bibitem{lee2021determining}
Jinhyuk Lee, Chao Wang, Rahul Malik, Yanhao Dong, Yimeng Huang, Dong-Hwa Seo,
  and Ju~Li.
\newblock Determining the criticality of li-excess for disordered-rocksalt
  li-ion battery cathodes.
\newblock {\em Advanced Energy Materials}, 11(24):2100204, 2021.

\bibitem{fong2024redox}
Richie Fong, Nauman Mubarak, Sang-Wook Park, Gregory Lazaris, Yiwei Liu, Rahul
  Malik, Dong-Hwa Seo, and Jinhyuk Lee.
\newblock Redox engineering of fe-rich disordered rock-salt li-ion cathode
  materials.
\newblock {\em Advanced Energy Materials}, 14(22):2400402, 2024.

\bibitem{bonnick2018insights}
Patrick Bonnick, Lauren Blanc, Shahrzad~Hosseini Vajargah, Chang-Wook Lee,
  Xiaoqi Sun, Mahalingam Balasubramanian, and Linda~F Nazar.
\newblock Insights into mg2+ intercalation in a zero-strain material:
  thiospinel mg x zr2s4.
\newblock {\em Chemistry of Materials}, 30(14):4683--4693, 2018.

\bibitem{lu2021searching}
Wang Lu, Juefan Wang, Gopalakrishnan Sai~Gautam, and Pieremanuele Canepa.
\newblock Searching ternary oxides and chalcogenides as positive electrodes for
  calcium batteries.
\newblock {\em Chemistry of Materials}, 33(14):5809--5821, 2021.

\bibitem{wang2015design}
Yan Wang, William~Davidson Richards, Shyue~Ping Ong, Lincoln~J Miara, Jae~Chul
  Kim, Yifei Mo, and Gerbrand Ceder.
\newblock Design principles for solid-state lithium superionic conductors.
\newblock {\em Nature materials}, 14(10):1026--1031, 2015.

\bibitem{canepa2017high}
Pieremanuele Canepa, Shou-Hang Bo, Gopalakrishnan Sai~Gautam, Baris Key,
  William~D Richards, Tan Shi, Yaosen Tian, Yan Wang, Juchuan Li, and Gerbrand
  Ceder.
\newblock High magnesium mobility in ternary spinel chalcogenides.
\newblock {\em Nature communications}, 8(1):1759, 2017.

\bibitem{jalem2014efficient}
Randy Jalem, Masanobu Nakayama, and Toshihiro Kasuga.
\newblock An efficient rule-based screening approach for discovering fast
  lithium ion conductors using density functional theory and artificial neural
  networks.
\newblock {\em Journal of Materials Chemistry A}, 2(3):720--734, 2014.

\bibitem{jalem2018bayesian}
Randy Jalem, Kenta Kanamori, Ichiro Takeuchi, Masanobu Nakayama, Hisatsugu
  Yamasaki, and Toshiya Saito.
\newblock Bayesian-driven first-principles calculations for accelerating
  exploration of fast ion conductors for rechargeable battery application.
\newblock {\em Scientific reports}, 8(1):5845, 2018.

\bibitem{sendek2017holistic}
Austin~D Sendek, Qian Yang, Ekin~D Cubuk, Karel-Alexander~N Duerloo, Yi~Cui,
  and Evan~J Reed.
\newblock Holistic computational structure screening of more than 12000
  candidates for solid lithium-ion conductor materials.
\newblock {\em Energy \& Environmental Science}, 10(1):306--320, 2017.

\bibitem{barsukov2012electrochemical}
Yevgen Barsukov, J~Ross Macdonald, et~al.
\newblock Electrochemical impedance spectroscopy.
\newblock {\em Characterization of materials}, 2:898--913, 2012.

\bibitem{itagaki2005licoo2}
Masayuki Itagaki, Nao Kobari, Sachiko Yotsuda, Kunihiro Watanabe, Shinichi
  Kinoshita, and Makoto Ue.
\newblock Licoo2 electrode/electrolyte interface of li-ion rechargeable
  batteries investigated by in situ electrochemical impedance spectroscopy.
\newblock {\em Journal of Power Sources}, 148:78--84, 2005.

\bibitem{adams2006bond}
Stefan Adams.
\newblock From bond valence maps to energy landscapes for mobile ions in
  ion-conducting solids.
\newblock {\em Solid State Ionics}, 177(19-25):1625--1630, 2006.

\bibitem{brown2009recent}
Ian~David Brown.
\newblock Recent developments in the methods and applications of the bond
  valence model.
\newblock {\em Chemical reviews}, 109(12):6858--6919, 2009.

\bibitem{hohenberg1964inhomogeneous}
Pierre Hohenberg and Walter Kohn.
\newblock Inhomogeneous electron gas.
\newblock {\em Physical review}, 136(3B):B864, 1964.

\bibitem{kohn1965self}
Walter Kohn and Lu~Jeu Sham.
\newblock Self-consistent equations including exchange and correlation effects.
\newblock {\em Physical review}, 140(4A):A1133, 1965.

\bibitem{frenkel2002understanding}
Daan Frenkel and Berend Smit.
\newblock Understanding molecular simulation.
\newblock {\em Academic Press, San Diego,}, 2(2.2):2--5, 2002.

\bibitem{mo2012first}
Yifei Mo, Shyue~Ping Ong, and Gerbrand Ceder.
\newblock First principles study of the li10gep2s12 lithium super ionic
  conductor material.
\newblock {\em Chemistry of Materials}, 24(1):15--17, 2012.

\bibitem{car1985unified}
Richard Car and Mark Parrinello.
\newblock Unified approach for molecular dynamics and density-functional
  theory.
\newblock {\em Physical review letters}, 55(22):2471, 1985.

\bibitem{meutzner2019computational}
F~Meutzner, T~Nestler, M~Zschornak, P~Canepa, GS~Gautam, S~Leoni, S~Adams,
  T~Leisegang, VA~Blatov, and DC~Meyer.
\newblock Computational analysis and identification of battery materials.
\newblock {\em Physical Sciences Reviews}, 4(1):20180044, 2019.

\bibitem{nestler2019towards}
Tina Nestler, Stanislav Fedotov, Tilmann Leisegang, and Dirk~C Meyer.
\newblock Towards al3+ mobility in crystalline solids: critical review and
  analysis.
\newblock {\em Critical Reviews in Solid State and Materials Sciences},
  44(4):298--323, 2019.

\bibitem{jonsson1998nudged}
Hannes J{\'o}nsson, Greg Mills, and Karsten~W Jacobsen.
\newblock Nudged elastic band method for finding minimum energy paths of
  transitions.
\newblock In {\em Classical and quantum dynamics in condensed phase
  simulations}, pages 385--404. World Scientific, 1998.

\bibitem{henkelman2000climbing}
Graeme Henkelman, Blas~P Uberuaga, and Hannes J{\'o}nsson.
\newblock A climbing image nudged elastic band method for finding saddle points
  and minimum energy paths.
\newblock {\em The Journal of chemical physics}, 113(22):9901--9904, 2000.

\bibitem{devi2022effect}
Reshma Devi, Baltej Singh, Pieremanuele Canepa, and Gopalakrishnan Sai~Gautam.
\newblock Effect of exchange-correlation functionals on the estimation of
  migration barriers in battery materials.
\newblock {\em npj Computational Materials}, 8(1):160, 2022.

\bibitem{he2018statistical}
Xingfeng He, Yizhou Zhu, Alexander Epstein, and Yifei Mo.
\newblock Statistical variances of diffusional properties from ab initio
  molecular dynamics simulations.
\newblock {\em npj Computational Materials}, 4(1):18, 2018.

\bibitem{rong2016efficient}
Ziqin Rong, Daniil Kitchaev, Pieremanuele Canepa, Wenxuan Huang, and Gerbrand
  Ceder.
\newblock An efficient algorithm for finding the minimum energy path for cation
  migration in ionic materials.
\newblock {\em The Journal of chemical physics}, 145(7), 2016.

\bibitem{deringer2019machine}
Volker~L Deringer, Miguel~A Caro, and G{\'a}bor Cs{\'a}nyi.
\newblock Machine learning interatomic potentials as emerging tools for
  materials science.
\newblock {\em Advanced Materials}, 31(46):1902765, 2019.

\bibitem{choyal2024constructing}
Vijay Choyal, Nidhish Sagar, and Gopalakrishnan Sai~Gautam.
\newblock Constructing and evaluating machine-learned interatomic potentials
  for li-based disordered rocksalts.
\newblock {\em Journal of Chemical Theory and Computation}, 20(11):4844--4856,
  2024.

\bibitem{gartner2020signatures}
Thomas~E Gartner~III, Linfeng Zhang, Pablo~M Piaggi, Roberto Car, Athanassios~Z
  Panagiotopoulos, and Pablo~G Debenedetti.
\newblock Signatures of a liquid--liquid transition in an ab initio deep neural
  network model for water.
\newblock {\em Proceedings of the National Academy of Sciences},
  117(42):26040--26046, 2020.

\bibitem{achar2021efficiently}
Siddarth~K Achar, Linfeng Zhang, and J~Karl Johnson.
\newblock Efficiently trained deep learning potential for graphane.
\newblock {\em The Journal of Physical Chemistry C}, 125(27):14874--14882,
  2021.

\bibitem{batatia2022design}
Ilyes Batatia, Simon Batzner, D{\'a}vid~P{\'e}ter Kov{\'a}cs, Albert Musaelian,
  Gregor~NC Simm, Ralf Drautz, Christoph Ortner, Boris Kozinsky, and G{\'a}bor
  Cs{\'a}nyi.
\newblock The design space of e (3)-equivariant atom-centered interatomic
  potentials.
\newblock {\em arXiv preprint arXiv:2205.06643}, 2022.

\bibitem{batatia2023foundation}
Ilyes Batatia, Philipp Benner, Yuan Chiang, Alin~M Elena, D{\'a}vid~P
  Kov{\'a}cs, Janosh Riebesell, Xavier~R Advincula, Mark Asta, Matthew Avaylon,
  William~J Baldwin, et~al.
\newblock A foundation model for atomistic materials chemistry.
\newblock {\em arXiv preprint arXiv:2401.00096}, 2023.

\bibitem{rhodes2025orb}
Benjamin Rhodes, Sander Vandenhaute, Vaidotas {\v{S}}imkus, James Gin, Jonathan
  Godwin, Tim Duignan, and Mark Neumann.
\newblock Orb-v3: atomistic simulation at scale.
\newblock {\em arXiv preprint arXiv:2504.06231}, 2025.

\bibitem{deng2023chgnet}
Bowen Deng, Peichen Zhong, KyuJung Jun, Janosh Riebesell, Kevin Han,
  Christopher~J Bartel, and Gerbrand Ceder.
\newblock Chgnet as a pretrained universal neural network potential for
  charge-informed atomistic modelling.
\newblock {\em Nature Machine Intelligence}, 5(9):1031--1041, 2023.

\bibitem{chen2022universal}
Chi Chen and Shyue~Ping Ong.
\newblock A universal graph deep learning interatomic potential for the
  periodic table.
\newblock {\em Nature Computational Science}, 2(11):718--728, 2022.

\bibitem{kim_sevennet_mf_2024}
Jaesun Kim, Jisu Kim, Jaehoon Kim, Jiho Lee, Yutack Park, Youngho Kang, and
  Seungwu Han.
\newblock Data-efficient multifidelity training for high-fidelity machine
  learning interatomic potentials.
\newblock {\em J. Am. Chem. Soc.}, 147(1):1042--1054, 2024.

\bibitem{devi2024optimal}
Reshma Devi, Keith~T Butler, and Gopalakrishnan Sai~Gautam.
\newblock Optimal pre-train/fine-tune strategies for accurate material property
  predictions.
\newblock {\em npj Computational Materials}, 10(1):300, 2024.

\bibitem{perdew1996generalized}
John~P Perdew, Kieron Burke, and Matthias Ernzerhof.
\newblock Generalized gradient approximation made simple.
\newblock {\em Physical review letters}, 77(18):3865, 1996.

\bibitem{anisimov1991band}
Vladimir~I Anisimov, Jan Zaanen, and Ole~K Andersen.
\newblock Band theory and mott insulators: Hubbard u instead of stoner i.
\newblock {\em Physical Review B}, 44(3):943, 1991.

\bibitem{wangoxidation2006}
Lei Wang, Thomas Maxisch, and Gerbrand Ceder.
\newblock Oxidation energies of transition metal oxides within the {GGA}+
  {\textbackslash}{emphU} framework.
\newblock {\em Physical Review B}, 73(19):195107, 2006.
\newblock Publisher: {APS}.

\bibitem{jones1989density}
Robert~O Jones and Olle Gunnarsson.
\newblock The density functional formalism, its applications and prospects.
\newblock {\em Reviews of Modern Physics}, 61(3):689, 1989.

\bibitem{sun2015strongly}
Jianwei Sun, Adrienn Ruzsinszky, and John~P Perdew.
\newblock Strongly constrained and appropriately normed semilocal density
  functional.
\newblock {\em Physical review letters}, 115(3):036402, 2015.

\bibitem{sai2018evaluating}
Gopalakrishnan Sai~Gautam and Emily~A Carter.
\newblock Evaluating transition metal oxides within dft-scan and scan+ u
  frameworks for solar thermochemical applications.
\newblock {\em Physical Review Materials}, 2(9):095401, 2018.

\bibitem{hellenbrandt2004inorganic}
Mariette Hellenbrandt.
\newblock The inorganic crystal structure database (icsd)—present and future.
\newblock {\em Crystallography Reviews}, 10(1):17--22, 2004.

\bibitem{jain2013commentary}
Anubhav Jain, Shyue~Ping Ong, Geoffroy Hautier, Wei Chen, William~Davidson
  Richards, Stephen Dacek, Shreyas Cholia, Dan Gunter, David Skinner, Gerbrand
  Ceder, et~al.
\newblock Commentary: The materials project: A materials genome approach to
  accelerating materials innovation.
\newblock {\em APL materials}, 1(1), 2013.

\bibitem{chu2018predicting}
Iek-Heng Chu, Sayan Roychowdhury, Daehui Han, Anubhav Jain, and Shyue~Ping Ong.
\newblock Predicting the volumes of crystals.
\newblock {\em Computational Materials Science}, 146:184--192, 2018.

\bibitem{ong2013python}
Shyue~Ping Ong, William~Davidson Richards, Anubhav Jain, Geoffroy Hautier,
  Michael Kocher, Shreyas Cholia, Dan Gunter, Vincent~L Chevrier, Kristin~A
  Persson, and Gerbrand Ceder.
\newblock Python materials genomics (pymatgen): A robust, open-source python
  library for materials analysis.
\newblock {\em Computational Materials Science}, 68:314--319, 2013.

\bibitem{kresse1993ab}
Georg Kresse and JJPRB Hafner.
\newblock Ab initio molecular dynamics for open-shell transition metals.
\newblock {\em Physical Review B}, 48(17):13115, 1993.

\bibitem{kresse1996efficient}
Georg Kresse and J{\"u}rgen Furthm{\"u}ller.
\newblock Efficient iterative schemes for ab initio total-energy calculations
  using a plane-wave basis set.
\newblock {\em Physical review B}, 54(16):11169, 1996.

\bibitem{kresse1999ultrasoft}
Georg Kresse and Daniel Joubert.
\newblock From ultrasoft pseudopotentials to the projector augmented-wave
  method.
\newblock {\em Physical review b}, 59(3):1758, 1999.

\bibitem{van2001first}
A~Van~der Ven, G~Ceder, M~Asta, and PD~Tepesch.
\newblock First-principles theory of ionic diffusion with nondilute carriers.
\newblock {\em Physical Review B}, 64(18):184307, 2001.

\bibitem{larsen2017atomic}
Ask~Hjorth Larsen, Jens~J{\o}rgen Mortensen, Jakob Blomqvist, Ivano~E Castelli,
  Rune Christensen, Marcin Du{\l}ak, Jesper Friis, Michael~N Groves, Bj{\o}rk
  Hammer, Cory Hargus, et~al.
\newblock The atomic simulation environment—a python library for working with
  atoms.
\newblock {\em Journal of Physics: Condensed Matter}, 29(27):273002, 2017.

\bibitem{nocedal1980updating}
Jorge Nocedal.
\newblock Updating quasi-newton matrices with limited storage.
\newblock {\em Mathematics of computation}, 35(151):773--782, 1980.

\bibitem{tekliye2022exploration}
Dereje~Bekele Tekliye, Ankit Kumar, Xie Weihang, Thelakkattu~Devassy Mercy,
  Pieremanuele Canepa, and Gopalakrishnan Sai~Gautam.
\newblock Exploration of nasicon frameworks as calcium-ion battery electrodes.
\newblock {\em Chemistry of Materials}, 34(22):10133--10143, 2022.

\bibitem{tekliye2024fluoride}
Dereje~Bekele Tekliye and Gopalakrishnan~Sai Gautam.
\newblock Fluoride frameworks as potential calcium battery cathodes.
\newblock {\em Journal of Materials Chemistry A}, 12(30):18993--19007, 2024.

\bibitem{deb2024exploration}
Debolina Deb and Gopalakrishnan Sai~Gautam.
\newblock Exploration of oxyfluoride frameworks as na-ion cathodes.
\newblock {\em Chemistry of Materials}, 36(24):11892--11904, 2024.

\bibitem{gautam2015first}
Gopalakrishnan~Sai Gautam, Pieremanuele Canepa, Rahul Malik, Miao Liu, Kristin
  Persson, and Gerbrand Ceder.
\newblock First-principles evaluation of multi-valent cation insertion into
  orthorhombic v 2 o 5.
\newblock {\em Chemical Communications}, 51(71):13619--13622, 2015.

\bibitem{gautam2016impact}
Gopalakrishnan~Sai Gautam, Xiaoqi Sun, Victor Duffort, Linda~F Nazar, and
  Gerbrand Ceder.
\newblock Impact of intermediate sites on bulk diffusion barriers: Mg
  intercalation in mg 2 mo 3 o 8.
\newblock {\em Journal of Materials Chemistry A}, 4(45):17643--17648, 2016.

\end{thebibliography}

\end{document}